\documentclass[%
amsmath,amssymb,
aps,
pre,
]{revtex4-2}

\usepackage{graphicx}% Include figure files
\usepackage{dcolumn}% Align table columns on decimal point
\usepackage{bm}% bold math
\usepackage{color}
\definecolor{darkblue}{rgb}{0,0,0.75}

\usepackage{lineno}
\begin{document}
	
	\preprint{APS/123-QED}

	\title{Dominant strategy in repeated games on networks}
	% Force line breaks with \\
	% \thanks{
		%     % \begin{enumerate}
			%     %     % \centering
			%     %     \item Center for Systems and Control, College of Engineering, Peking University, Beijing 100871, China
			%     %     \item School of Automation, Beijing Institute of Technology, Beijing 100081, China
			%     %     \item Center for Multi-Agent Research, Institute for Artificial Intelligence, Peking University, Beijing 100871, China
			%     %     \item [$\dag$] These authors contributed equally to this work
			%     %     \item [$*$] Corresponding author. E-mail: amingli@pku.edu.cn
			%     % \end{enumerate}
		%     }%
	
	\author{Xiaochen Wang$^{1}$ and Aming Li$^{1,2,*}$
		\\
		% \centering
		\footnotesize{$^{1}$Center for Systems and Control, College of Engineering, Peking University, Beijing 100871, China}\\
		\footnotesize{$^{2}$Center for Multi-Agent Research, Institute for Artificial Intelligence, Peking University, Beijing 100871, China}\\
		\footnotesize{$*$ Corresponding author. E-mail: amingli@pku.edu.cn}\\
	}

\begin{abstract}
	Direct reciprocity, stemming from repeated interactions among players, is one of the fundamental mechanisms for understanding the evolution of cooperation. 
	However, canonical strategies for the repeated prisoner's dilemma, such as Win-Stay-Lose-Shift and Tit-for-Tat, fail to consistently dominate alternative strategies during evolution. 
	This complexity intensifies with the introduction of spatial structure or network behind individual interactions, where nodes represent players and edges represent their interactions.
	Here, we propose a new strategy, ``Cooperate-Stay-Defect-Tolerate" (CSDT), which can dominate other strategies within networked populations by adhering to three essential characteristics. 
	This strategy maintains current behaviour when the opponent cooperates and tolerates defection to a limited extent when the opponent defects.
	We demonstrate that the limit of tolerance of CSDT can vary with the network structure, evolutionary dynamics, and game payoffs. 
	Furthermore, we find that incorporating the Always Defect strategy (ALLD) can enhance the evolution of CSDT and eliminate strategies that are vulnerable to defection in the population, providing a new interpretation of the role of ALLD in direct reciprocity.
	Our findings offer a novel perspective on how cooperative strategy evolves on networked populations.
\end{abstract}
	\maketitle
	%----------------------------------------------------------------------------------------
	
	\section{Introduction}
%   Following the inception of game theory by von Neumann and Nash \cite{morgenstern1953theory,nash1950equilibrium}, and the subsequent experimentation conducted by Flood and Dresher \cite{flood1952some}, the prisoner's dilemma has endured as a unifying paradigm, illustrating the concepts of direct reciprocity and social dilemma for decades.
%   Consider two players playing a game under the following framework:
%   \begin{equation*}
	%       \bordermatrix{%
		%           & \text{C} & \text{D} \cr
		%           \text{C} & R & S \cr
		%           \text{D} & T & P \cr
		%       }.
	%   \end{equation*}
%   Each player can choose to cooperate (C) or defect (D).
%   A prisoner's dilemma is defined by the conditions $T>R>P>S$ and $2R>T+S$.
%   In this context, defection becomes the rational choice for players to maximize their payoffs. 
%   As a result, the Nash equilibrium is for both players to defect, deviating from the Pareto-optimal outcome represented by mutual cooperation.
%   This situation highlights a conflict between collective interests and player decision-making.

The prisoner's dilemma illustrates a conflict where the Nash equilibrium \cite{nash1950equilibrium} (mutual defection) deviates from the Pareto optimum (mutual cooperation).
However, not only in reality but even in the Flood and Dresher's experiment \cite{flood1952some} wherein this model was introduced, mutual defection did not manifest as strongly as anticipated. 
Nash attributed this phenomenon to the repeated nature of the game \cite{flood1958some}, which has become a pivotal explanation for the prevalence of collective cooperation, known as direct reciprocity \cite{nowak2006five}.
Classic strategies of repeated games, such as Always Cooperate (ALLC) and Always Defect (ALLD), consistently repeat the same actions (Fig.~1a). 
Tit-for-Tat (TFT) \cite{axelrod1981evolution} reciprocates the opponent's last move and is remarkably successful in promoting cooperation without the presence of noise. 
However, if a player defects mistakenly, two players employing the TFT strategy will inevitably encounter at least one defector, thereby making mutual cooperation vulnerable to errors.
Generous Tit-for-Tat (GTFT) \cite{nowak1992TFT} and Win-Stay Lose-Shift (WSLS) \cite{nowak1993WSLS} address this issue by maintaining a certain probability of cooperation in the face of defection.
Zero-Determinant (ZD) strategies \cite{press2012iterated} allow players to exert direct control over their opponent's payoff. 
And partner and rival strategies \cite{hilbe2015partners,hilbe2018partners} further expand the strategic landscape by imposing constraints on the payoffs of opponents.

Despite their favourable attributes in pairwise interactions, none of these strategies consistently dominate other strategies across all evolutionary scenarios \cite{nowak1992TFT,nowak1993WSLS,adami2013evolutionary,stewart2013from,hilbe2013evolution,szolnoki2014defection}. 
Indeed, they are particularly susceptible to invasion by ALLD strategies, especially when the temptation to defect is high \cite{nowak1992TFT,nowak1993WSLS}. 
However, the prevalence of cooperative behaviour in real-world scenarios underscores the need for strategies that can evolve robustly. 
Expanding on the idea of the ``Good strategy" \cite{akin2015what}, which is an extension of the Zero-Determinant (ZD) strategy designed to secure mutual cooperation, subsequent research has introduced strategies that blend the principles of goodness with robustness \cite{stewart2013from}. 
However, these conclusions are constrained within the ZD framework and are primarily applicable to well-mixed populations.
%	To comprehensively investigate dominant strategies across diverse populations (Fig.~1b), it is crucial to explore a broader strategy space and more varied population structures, transcending existing limitations.

Real-world interactions extend beyond unstructured, well-mixed populations. Network representations offer a powerful framework for modelling diverse population structures, where nodes represent players, and edges represent their interactions \cite{barabasi1999emergence,watts1998collective, allen2017evolutionary}. By constraining patterns of interaction and dispersal, networked populations promote the evolution of cooperative behavior \cite{ohtsuki2006simple,allen2017evolutionary,tarnita2009set,nowak1992evolutionary,li2020evolution,tarnita2009strategy,tarnita2010evolutionary}, which is referred to the network reciprocity \cite{nowak2006five}. 
However, previous research on repeated game strategies largely neglected the influence of network structure, resulting in classical strategies failing to evolve in more complex, real-world populations. 
This oversight creates a gap between the mechanisms of direct reciprocity and network reciprocity, which raises an intriguing question: Is there a dominant strategy capable of prevailing over others across any network?

To explore a wide range of populations and a broader strategy space (Fig.~1b), we systematically analyse the nature of dominant strategies and identify a type of strategy that can prevail over others, regardless of the number and types of opponents. 
We term this new strategy ``Cooperate-Stay-Defect-Tolerate" (CSDT), which maintains the current move when the opponent cooperates and tolerates defection. 
The composition of these strategies depends on environmental factors such as network structure and game payoffs. 
However, CSDT will always exist and dominate across any network structure.
Surprisingly, contrary to previous beliefs, we find that ALLD can enhance the evolutionary advantage of CSDT and improve the population's resilience against defection.
Our results unveil the complexity of integrating direct reciprocity with network reciprocity, offering a potential explanation for the existence and prevalence of cooperative behaviour.

%------------------------------------------------------------------------
%   \end{spacing}
%\clearpage
\section{Model}
Consider two players playing the following game:
\begin{equation}
\begin{aligned}
	&~~~~\text{Player B}\\
	\text{Player A}~~&
	\bordermatrix{%
		& \text{C} & \text{D} \cr
		\text{C} & R & S \cr
		\text{D} & T & P \cr
	},
\end{aligned}
\label{payoffmat}
\end{equation}
where each entry indicates the payoff of player A, who is playing against another player B.
Each player may choose to cooperate (C) or defect (D), and the game corresponds to the prisoner’s dilemma when $T>R>P>S$ and $2R>T+S$.
From a perspective of public interest, it is beneficial for all players to cooperate, representing the Pareto optimum \cite{osborne2004introduction}.
However, regardless of the opponent's choice, a player can maximize its own payoff unilaterally by defection. 
Consequently, defection for each player is the only Nash equilibrium.

\begin{center}
\begin{figure}[]
	\centering
	\includegraphics[width=1\textwidth]{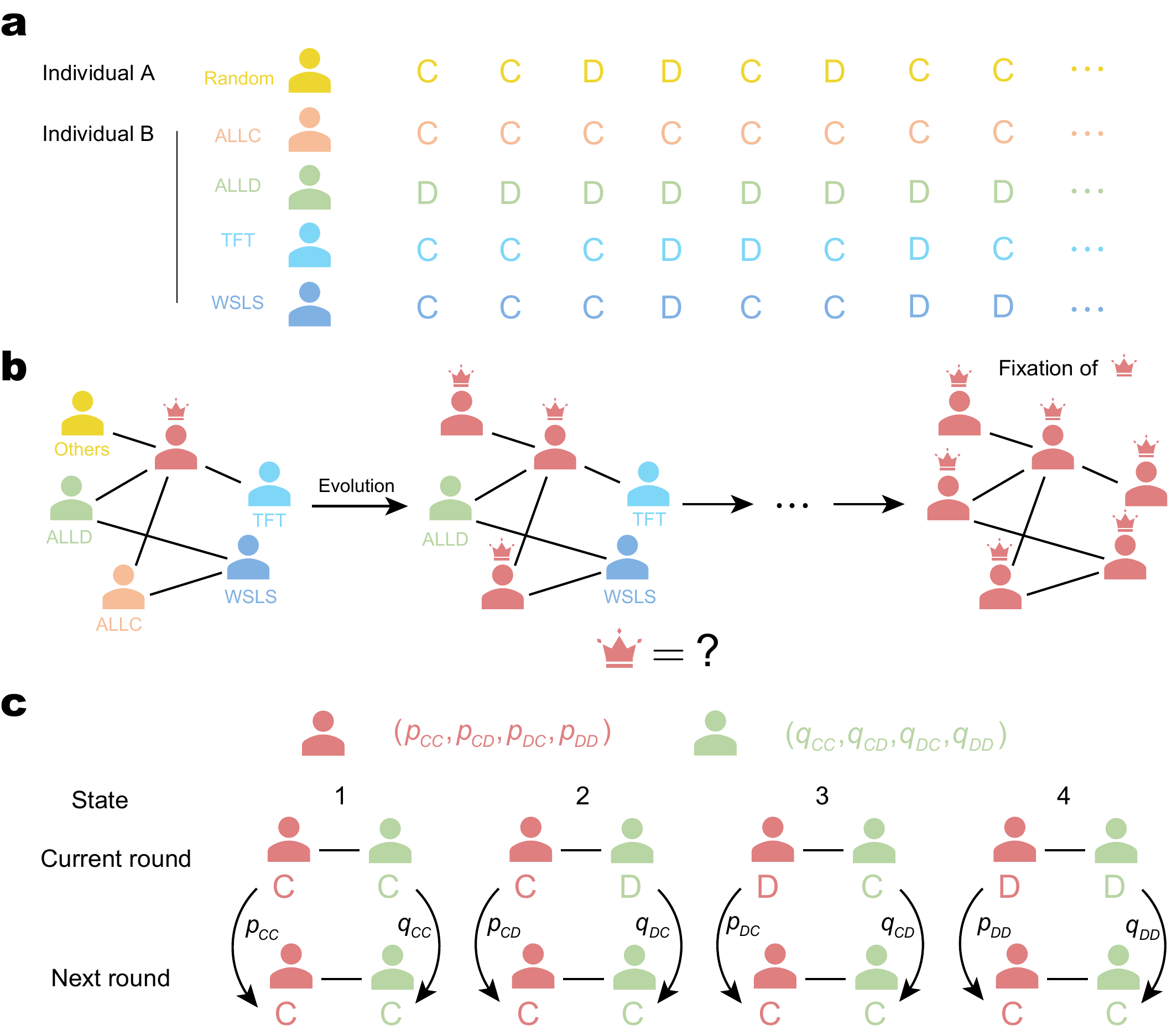}
	
	\caption{
		\textbf{Direct reciprocity strategies that dominate networked populations.}
		\textbf{a},
		Illustration on the classical strategies for repeated prisoner's dilemma between individuals A and B, and here we list the strategy: Always Cooperate (ALLC), ALways Defect (ALLD), Tit-For-Tat (TFT), and Win-Stay-Lose-Shift (WSLS).
		\textbf{b},
		The interconnections among players within the population can be delineated as a network structure. 
		Each player can adopt distinct strategies. 
		Here, we investigate which specific strategies are more likely to dominate other strategies during evolution.
		\textbf{c}, 
		We focus on the ``memory-one" strategy, represented as a vector with four elements: ($p_{\text{CC}},p_{\text{CD}},p_{\text{DC}},p_{\text{DD}}$). 
		Each element indicates the probability of a player employing this strategy to cooperate in the next round following specific game outcomes. 
	}
	\label{fig: 1}
\end{figure}
\end{center}

The repeated prisoner's dilemma implies that players can determine their next actions based on the memory of previous rounds of the game.
The basis upon which each player assesses whether to cooperate or defect in the next round is referred to as a strategy.
Here, we focus on the memory-one strategies \cite{press2012iterated}, characterized by utilizing solely the actions of the player and opponent in the immediate prior round of the game. 
The memory-one strategy takes the form $\bm{p}=(p_{\text{CC}},p_{\text{CD}},p_{\text{DC}},p_{\text{DD}})$, where each component represents the probability of cooperating in the next round given the outcome of the current round (Fig.~\ref{fig: 1}c). 
For example, a player cooperates with probability $p_{\text{CD}}$ when she cooperated while the opponent defected in the last round (CD).
Consider an alternate player employing the memory-one strategy $\bm{q}=(q_{\text{CC}},q_{\text{CD}},q_{\text{DC}},q_{\text{DD}})$. 
The repeated game between the two players can be conceived as a Markov process, encapsulated by a Markov matrix:
\begin{equation}
\label{Markov}
\bm{M}=
\begin{bmatrix}
	p_{\text{CC}}q_{\text{CC}} & p_{\text{CC}}(1-q_{\text{CC}}) & (1-p_{\text{CC}})q_{\text{CC}} & (1-p_{\text{CC}})(1-q_{\text{CC}})\\
	p_{\text{CD}}q_{\text{DC}} & p_{\text{CD}}(1-q_{\text{DC}}) & (1-p_{\text{CD}})q_{\text{DC}} & (1-p_{\text{CD}})(1-q_{\text{DC}})\\
	p_{\text{DC}}q_{\text{CD}} & p_{\text{DC}}(1-q_{\text{CD}}) & (1-p_{\text{DC}})q_{\text{CD}} & (1-p_{\text{DC}})(1-q_{\text{CD}})\\
	p_{\text{DD}}q_{\text{DD}} & p_{\text{DD}}(1-q_{\text{DD}}) & (1-p_{\text{DD}})q_{\text{DD}} & (1-p_{\text{DD}})(1-q_{\text{DD}})
\end{bmatrix}.
\end{equation}
Memory-one strategies have been extensively used in classical and evolutionary game theory.
Their simple forms encode many interesting and intricate human behaviours.
One might think that a longer memory would provide a greater advantage in the game. 
However, researchers have shown that when the same game is infinitely repeated, longer memory does not provide an advantage in terms of payoff \cite{press2012iterated}. 

From the perspective of evolutionary game theory, strategies undergo evolution within a population of fixed size $N$, where each player may adopt a distinct strategy from $n$ strategies. Following interactions, player $i$'s payoff, denoted as $s_i$, is translated into fitness using the transformation $F_i = e^{\omega s_i}$, with $\omega$ signifying the selection intensity \cite{nowak2006five}. 
In line with previous studies, here we focus on the situations where $\omega$ approaches 0, indicating that selection operates weakly, and game payoffs play a slight role in strategy updates. 
For strategy updating, players update their strategies based on specific rules linked to their fitness.
Without loss of generality, we consider the canonical death-birth (DB) rule \cite{ohtsuki2006simple}, where a player $i$ is randomly selected to update its strategy, and it forgoes its own strategy and imitates one of its neighbours $j$ with a probability proportional to its fitness $F_j/\sum_{k\in\Omega_i}F_k$, where $\Omega_i$ is the set of $i$'s neighbours.
Based on such evolutionary dynamics, we investigate the fixation probability of strategy X, $\rho_{X}$, which represents the probability that strategy X takes over the whole population where initially each strategy is used by $N/n$ players. 
If strategy X has the highest fixation probability, it is considered to dominate other strategies.

The concept of the ``structure coefficient" provides a robust framework for analysing evolutionary dynamics and fixation probabilities within diverse structured populations \cite{tarnita2009strategy,tarnita2010evolutionary,tarnita2011multiple}. 
The dominance of Strategy X over Strategy Y is determined by the inequality $\sigma s_{xx} + s_{xy} \ge \sigma s_{yy} + s_{yx}$, where $s_{xy}$ represents the payoff of Strategy X against Strategy Y, and $\sigma$ denotes the structure coefficient. 
This coefficient is influenced by model specifics and evolutionary dynamics, including population structure and update rules. 
A larger $\sigma$ corresponds to an environment where cooperation is more likely to evolve.
This powerful tool enables us to integrate network reciprocity into the study of direct reciprocity.

%——————————————————————————————————————————————————————————

\section{Results}
\subsection{Characteristics required for the dominant strategy}

We now investigate the specific form of the dominant strategy. 
A dominant strategy is one that never has an evolutionary disadvantage compared to any other strategy. Mathematically, this means that for a strategy X, the inequality $\sigma (s_{xx} - s_{yy}) - (s_{yx} - s_{xy}) \ge 0$ must hold true regardless of the opponent's strategy Y \cite{tarnita2009strategy}. 
Upon closer examination of the given conditions, it becomes apparent that $s_{yy}$ depends on the opponent's strategy type, satisfying $P \le s_{yy} \le R$. 
For strategy X to meet this criterion, adjustments must be made to the other three terms in the inequality. 
Consequently, we decompose this problem into the following three aspects.

\begin{center}
\begin{figure}[h]
	\centering
	\includegraphics[width=1\textwidth]{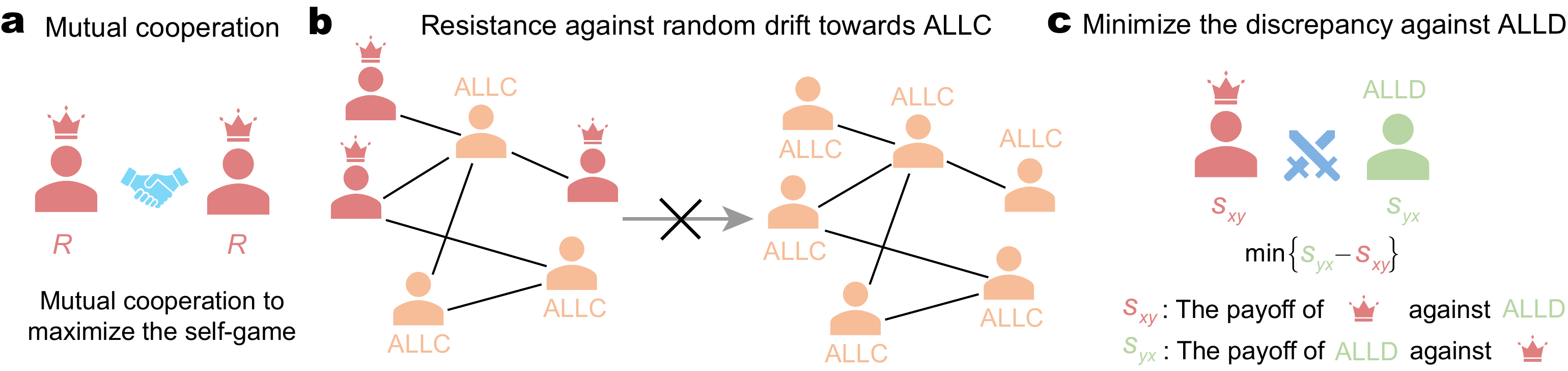}
	
	\caption{
		\textbf{Characteristics required for strategies to dominate a population.}
		\textbf{a},
		An effective strategy must excel in maximizing its payoff when encountering an identical strategy. In the context of the repeated prisoner's dilemma game, this effectiveness is encapsulated by the reward $R$, symbolizing the mutual cooperation between both participants.
		\textbf{b},
		Furthermore, the strategy should be resilient against random shifts favouring ALLC. 
		In scenarios with only ALLC and this strategy, it should prevent being supplanted by ALLC, as ALLC cannot withstand invasion by ALLD.
		\textbf{c},
		The last characteristic entails that if this strategy produces lower payoffs when interacting with other strategies, it should aim to reduce this discrepancy. 
		Computational analysis reveals that ALLD exploits this strategy to the maximum degree. 
		Consequently, the focus shifts to narrowing the payoff differential in interactions with ALLD.
	}
	\label{fig: 2}
\end{figure}
\end{center}

The structure coefficient $\sigma$ is influenced by the population structure and evolutionary dynamics (update rules) under weak selection and can be applied to the payoff structure of any pairwise game as indicated in Eq.~(\ref{payoffmat}). 
This conclusion is applicable to any network and dynamics discussed, thus encompassing all relevant structural aspects. 
The coefficient $\sigma$ essentially acts as a weight on the payoffs for interactions of the same type, illustrating the phenomenon of aggregation. The aggregation of the same strategy is crucial for a strategy to achieve an evolutionary advantage, particularly in structured populations \cite{nowak2006five,tarnita2010evolutionary,allen2017evolutionary}.
Consequently, players using a dominant strategy need to maximise their payoffs when interacting with others employing the same strategy. 
In the repeated prisoner's dilemma, the highest achievable payoff when identical strategies interact is the payoff for mutual cooperation, denoted as $R$ (Fig.~\ref{fig: 2}a).
Thus, for the dominant strategy X, the payoff against players using the same strategy must be $s_{xx} = R$. 
Solving this requirement and combining it with Eq.~(\ref{Markov}) (see Supplementary Note 3), we obtain
\begin{subequations}
\begin{align}
	&p_{\text{CC}}=1,\label{pcc1}
	\\
	&p_{\text{DD}}\neq0,\label{pdd0}
	\\
	&\neg(p_{\text{CD}}=1\land p_{\text{DC}}=0). \label{pcd_pdc}
\end{align}
\end{subequations}
Here, $\neg(A \land B)$ denotes that $A$ and $B$ cannot both be true simultaneously. 
Mathematically, these conditions ensure that the stationary distribution is unique and equal to $(1, 0, 0, 0)$. 
Intuitively, Eq.~(\ref{pcc1}) represents the sustainability of cooperation. 
When both players use the dominant strategy to cooperate, it ensures the continuation of cooperation in subsequent rounds. 
Conversely, the other two conditions, Eq.~(\ref{pdd0}) and Eq.~(\ref{pcd_pdc}), ensure that the dynamics do not become locked into other states besides CC. 
For example, if $p_{\text{DD}} = 0$, both players will continue to defect after an initial defection and will never return to cooperation. 
This vulnerability is precisely the Achilles' heel of the famous TFT, whereas the WSLS strategy effectively avoids this issue. 
Similarly, if $p_{\text{CD}} = 1$ and $p_{\text{DC}} = 0$ hold simultaneously, the game state may become deadlocked, with one player always cooperating while the other always defects (CD or DC).

However, if the dominant strategy unconditionally cooperates, it cannot dominate ALLC, as both strategies yield a payoff of $R$ after interactions. 
This vulnerability is critical because it implies that the strategy could be supplanted by ALLC, leaving the population susceptible to invasion by ALLD. 
Therefore, an additional essential characteristic of a dominant strategy is its ability to prevent a neutral drift towards ALLC (Fig.~\ref{fig: 2}b). 
This leads to
\begin{equation}
p_{\text{DC}}=0, \quad p_{\text{CD}}\neq 1.
\label{pdc0}
\end{equation}
Equation~(\ref{pdc0}) indicates that dominant strategies can exploit the ``kindness" of their opponents. 
Once a player realises that the opponent will always choose to cooperate, the player will continually defect to maximise their own payoffs.

Easy to prove, however, if one wants to ensure that her opponent's payoff is never greater than hers, i.e., $s_{yx}<s_{xy}$, she must have $p_{\text{DD}}=0$.
However, as previously analysed, this condition contradicts the requirements for a dominant strategy to sustain long-lasting cooperation (Eq.~(\ref{pdc0})). 
Consequently, the dominant strategies inevitably face disadvantages when confronting certain strategies. 
To address this disadvantage, a trade-off is necessary where the benefits of aggregation outweigh the payoff discrepancies. 
This results in a maximum-minimisation problem, where we aim to minimise the discrepancy against other strategies to identify the dominant strategy, which can be expressed as 
\begin{equation}
\min_{\bm{p}}\left\{\max_{\bm{q}}\left\{ s_{yx} - s_{xy} \right\}\right\}.
\end{equation}

The feasible region of strategy payoffs can be determined by their interactions with eleven pure strategies \cite{mcavoy2019reactive}, which allows for the analysis of the properties of any strategy based on this result (See Supplementary Note 2).
Combining Eqs.~(\ref{pcc1}) and (\ref{pdc0}), we find that the maximum of the payoff difference arises from interactions with ALLD.
Thus, the problem becomes to minimize the discrepancy against ALLD (Fig.~\ref{fig: 2}c), which generates
\begin{equation}\label{DefeatALLD}
\sigma(R-P)\ge(s_{yx}-s_{xy}),
\end{equation}
where Y is ALLD.
Cooperative strategies are clearly at a disadvantage in repeated games when facing ALLD ($s_{xy} < s_{yx}$). 
However, as Eq.~(\ref{DefeatALLD}) shows, due to the influence of the structure coefficient $\sigma$, which represents aggregations in structured populations, ALLD cannot gain an evolutionary advantage over the strategy when the payoff difference is sufficiently small. 
In such cases, the strategy can become dominant in the evolutionary process.
To satisfy the condition for the dominant strategy to win over ALLD (Eq.~\ref{DefeatALLD}), we have
\begin{equation}
\sigma (R-P)+\frac{(T-S) p_{\text{DD}}}{p_{\text{CD}}-1-p_{\text{DD}}}>0.
\label{ResiALLD}
\end{equation}
We find that the condition in Eq.~(\ref{ResiALLD}) depends on both the payoff and the structure coefficient, which represents the environment of the evolutionary game.

\subsection{Feasible region of dominant strategy}
Combining Eqs.~(\ref{pcc1}-c), (\ref{pdc0}), and (\ref{ResiALLD}), we can deduce the form of a dominant strategy.
When $(T-S)>\sigma(R-P)$,
\begin{equation}
p_{\text{CC}}=1,\quad p_{\text{DC}}=0,\quad0<p_{\text{DD}}<\frac{\sigma(R-P)}{(T-S)-\sigma(R-P)}(1-p_{\text{CD}}).
\label{CSDT2}
\end{equation} 
And when $(T-S)\le\sigma(R-P)$, 
\begin{equation}
p_{\text{CC}}=1,\quad0\le p_{\text{CD}}<1,\quad p_{\text{DC}}=0,\quad 0<p_{\text{DD}}\le1.
\label{CSDT1}
\end{equation}
All strategies in this set are dominant under the corresponding dynamics, population structure, and payoff matrix. 
Notably, when any two strategies within this set interact, their payoffs are always $R$, indicating that no strategy in the set gains an advantage over another. 
However, these strategies do hold an evolutionary advantage when facing those outside the set. 
We observe that players tend to maintain their actions when opponents cooperate and exhibit a degree of tolerance, albeit limited, when facing defections. 
Based on these characteristics, we refer to these new strategies as ``Cooperate-Stay-Defect-Tolerate" (CSDT), identifying them as the dominant strategies we seek.

In fact, $p_{\text{CD}}$ and $p_{\text{DD}}$ represent the probabilities that players will cooperate after their opponents defect, reflecting the strategy's tolerance. 
Equations~(\ref{CSDT2}) and (\ref{CSDT1}) describe the tolerance limits for CSDT under different environments, depending on the interaction between the payoff and structure coefficient. 
When the environment is particularly harsh—meaning there are greater disadvantages when facing defection (e.g., ALLD) such that $(T-S) > \sigma(R-P)$—the tolerance of this strategy is constrained (Eq.~(\ref{CSDT2})).
In a relatively benign environment (i.e., less disadvantageous when facing ALLD, $(T-S)\le\sigma(R-P)$), tolerance can be quite high (Eq.~(\ref{CSDT1})). 
In fact, in such a mild environment, even ALLC can prevail over ALLD \cite{ohtsuki2006simple}. 
In increasingly harsh environments, strategies with higher tolerance levels are more likely to be eliminated. 
Beyond the limit of tolerance, these strategies no longer belong to CSDT and, therefore, cannot dominate other strategies.

In Supplementary Note 3.3, we provide a unified form of these two equations (Eq.~(S6)). We define the tolerance limit as
\begin{equation}
L = \arctan\left(\frac{\sigma(P-R)}{(T-S) + \sigma(P-R)}\right).
\end{equation}
The feasible region for CSDT on the $p_{\text{CD}} - p_{\text{DD}}$ plane is determined by the intersection of two areas: $L < \theta \le \pi$, where $\tan(\theta) = \frac{p_{\text{DD}}}{p_{\text{CD}} - 1}$, and the square defined by $0 \le p_{\text{CD}} < 1$ and $0 < p_{\text{DD}} \le 1$. 
A larger $L$ corresponds to a smaller feasible region, indicating a lower tolerance for defection (Fig.~S2). 
As shown in Eq.~(\ref{CSDT2}), there is a trade-off between $p_{\text{CD}}$ and $p_{\text{DD}}$: if a strategy exhibits high tolerance for defection in one scenario, it must reduce its tolerance in the other. 

\begin{center}
\begin{figure}[h]
	\centering
	\includegraphics[width=1\textwidth]{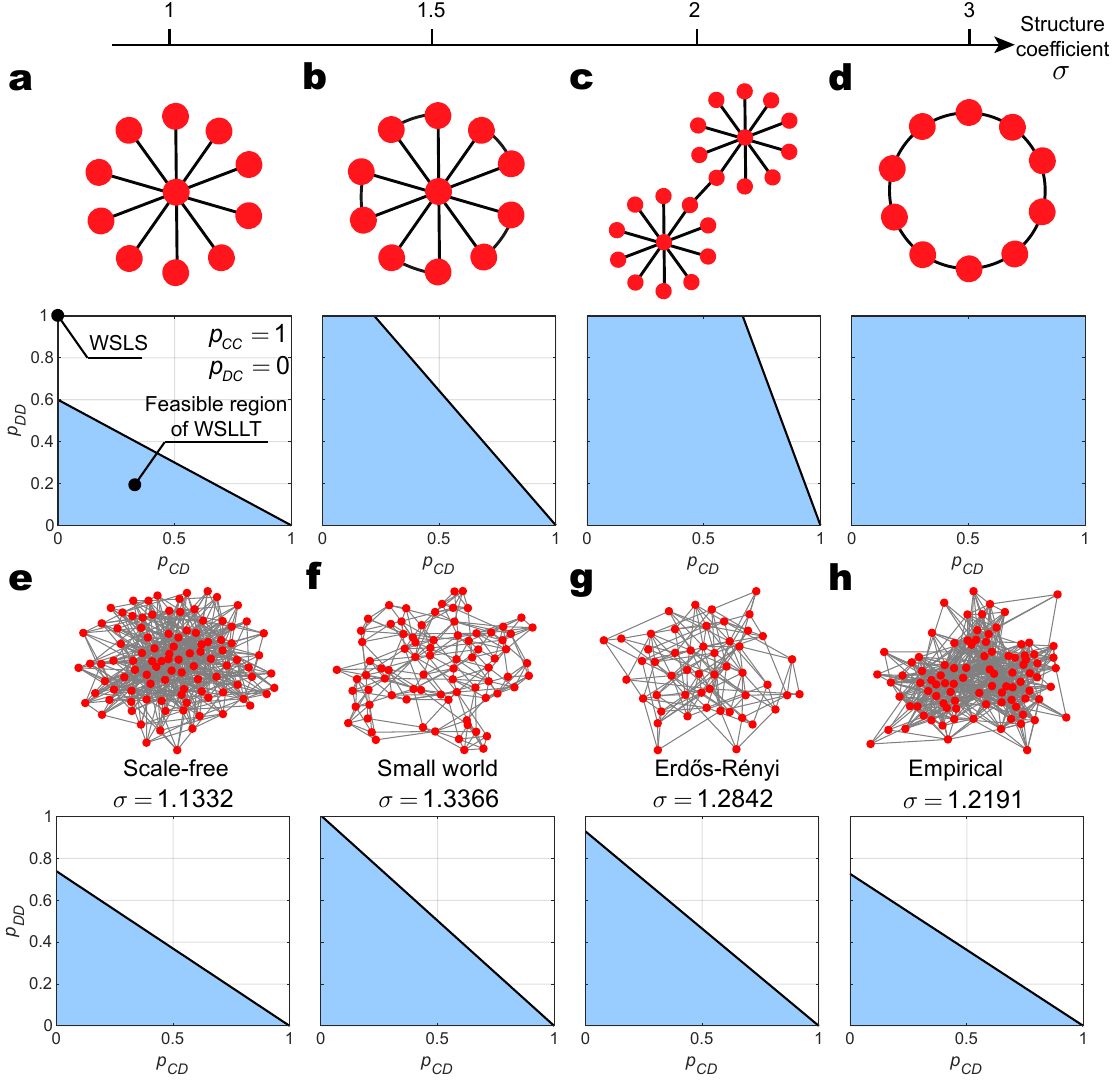}
	
	\caption{
		\textbf{Feasible regions of CSDT for different networks.}
		We plot the feasible region of CSDT on various networks with different structure coefficients $\sigma$.
		\textbf{a},
		For stars with a large population size $N$ and $\sigma=1$, the feasible region is the smallest.
		\textbf{b},
		A ceiling fan with $\sigma=3/2$ has a larger feasible region than stars. And WSLS is included in CSDT.
		\textbf{c},
		Connecting two identical stars via their leaves results in $\sigma=2$.
		\textbf{d},
		A circle with $\sigma=3$ yields $0 \le p_{\text{CD}} < 1$ and $0 < p_{\text{DD}} \le 1$, producing the largest feasible region.
		\textbf{e-h},
		we present examples of classic synthetic and empirical networks, showing how the feasible region varies with different structure coefficients.
		We have $T=6,R=3,P=1,S=-2$.
	}
	\label{fig: 3}
\end{figure}
\end{center}

Figures~\ref{fig: 3}a-d depict four distinct networks with varying structure coefficients. 
As the structure coefficient increases, indicating a milder environment, the feasible region expands accordingly. 
The canonical WSLS strategy can be included when the feasible region is sufficiently wide (Fig.~\ref{fig: 3}b). 
Figure~\ref{fig: 3}d illustrates a network structure with a high $\sigma$, where the feasible region of the dominant strategy reaches its maximum extent under a given payoff. 
We also plot the feasible regions on three synthetic networks and an empirical network, as shown in Figs.~\ref{fig: 3}e-h.
The range of CSDT varies significantly across different networks, underscoring the profound impact that network structure has on direct reciprocity.

Note that the lower bound of the structure coefficient is $1$, which corresponds to the most extreme conditions (Fig.~\ref{fig: 3}a) \cite{nowak2010evolutionary}. 
The feasible region for CSDT expands monotonically as the structure coefficient $\sigma$ increases. 
Consequently, we can define the kernel-CSDT by
\begin{equation}
p_{\text{CC}}=1,\quad p_{\text{DC}}=0,\quad0<p_{\text{DD}}<\frac{(R-P)}{(T-S)-(R-P)}(1-p_{\text{CD}}).
\label{kernel}
\end{equation}
Such strategies can dominate other strategies across any network structure. 
The kernel-CSDT is determined solely by the payoff and is guaranteed to be non-empty. 
This implies that there are always strategies capable of dominating the population, regardless of the network structure. 
Furthermore, in a mild environment (i.e., when $(T-S)$ is small), WSLS can be one of the kernel-CSDTs.

\subsection{The evolution of CSDT and the catalytic effect of ALLD}

\begin{center}
\begin{figure}[h]
	\centering
	\includegraphics[width=1\textwidth]{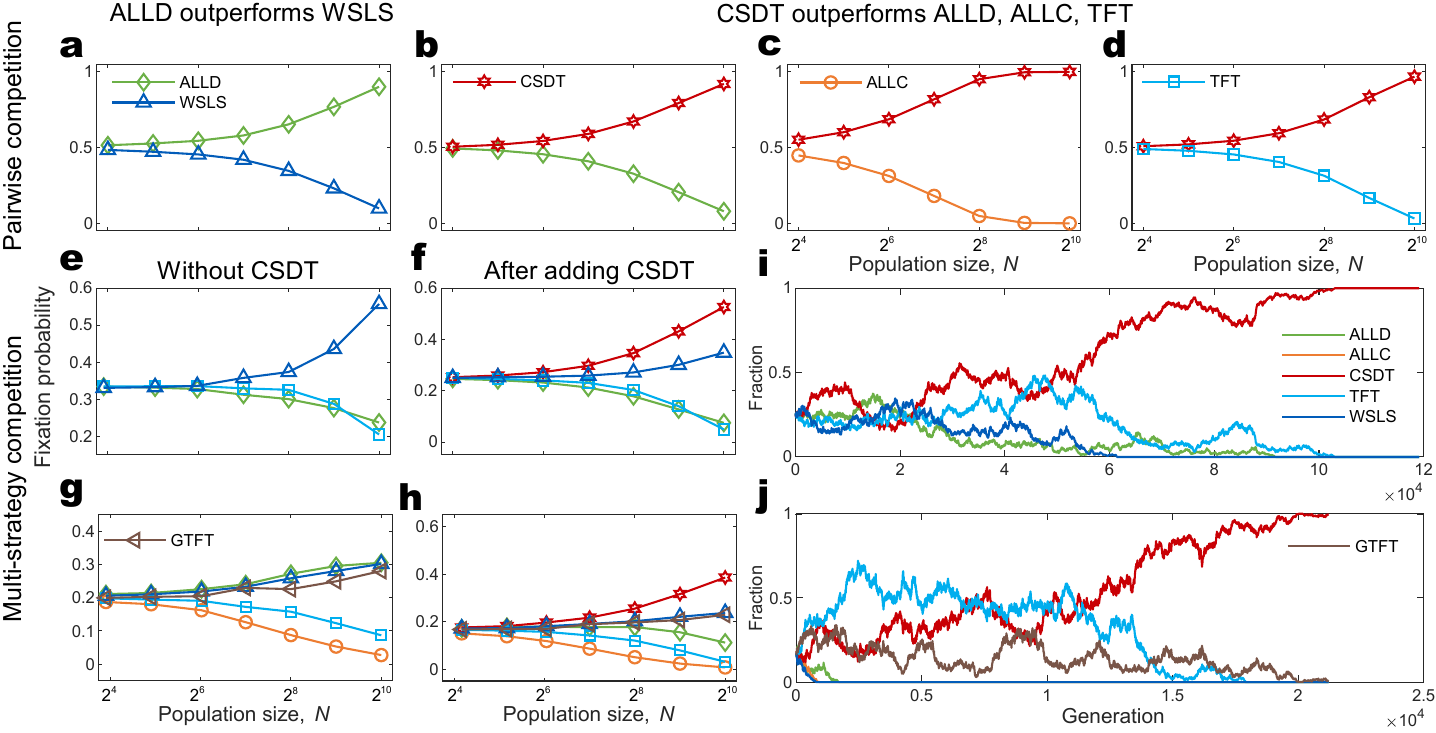}
	
	\caption{
		\textbf{CSDT outperforms other strategies.}
		We plot the fixation probabilities of different strategies as a function of the population size $N$.
		\textbf{a},
		When the population consists solely of ALLD and WSLS strategies, ALLD dominates the population.
		Furthermore, the advantage of ALLD increases with the augmentation of the population size $N$.
		\textbf{b},
		With only ALLD and CSDT present in the population, CSDT takes precedence.
		\textbf{c, d},
		CSDT can also dominate over TFT or ALLC.
		\textbf{e, f},
		When the population includes ALLD, TFT, and WSLS, the introduction of CSDT can also lead to its dominance within the population.
		\textbf{g, h},
		When the population includes ALLD, ALLC, TFT, WSLS, and GTFT, CSDT can also dominate other strategies.
		\textbf{i, j},
		We plot the change in fractions of different strategies over generations in a single round of evolution on the network with $N=500$. 
		Here, the population structure is modelled as a random-regular network with the degree $k=6$.
		We set $T=6,R=3,P=1,S=-2$, and $\omega=0.01$.
		The specific form of the CSDT is $(1, 0, 0, 0.2)$.
	}
	\label{fig: 4}
\end{figure}
\end{center}

To verify whether CSDT can dominate other strategies over evolution, we first conduct simulations on a $k$-regular network, where $k$ denotes the degree of each node. 
For such networks, the structure coefficient $\sigma$ is given by $\sigma=[(k+1)N-4k]/[(k-1)N]$.
We investigate a harsh environment where WSLS does not meet the criteria for CSDT, and ALLD prevails over WSLS (Fig.~\ref{fig: 4}a). 
Interestingly, we demonstrate that CSDT can outperform ALLD, ALLC, and TFT in evolutionary dynamics under pairwise competition (Figs.~\ref{fig: 4}b-d). 
This advantage becomes increasingly pronounced as the population size $N$ becomes larger.

When it comes to the evolution of multi-strategies, the condition for strategy X to dominate other strategies becomes
\begin{equation}
1/n\sum_{y=1}^{n}(\sigma s_{xx}+s_{xy}-s_{yx}-\sigma s_{yy})\ge0,
\label{multi}
\end{equation}
where $n$ is the number of strategies \cite{tarnita2011multiple}.
We have already demonstrated that CSDT holds for each pairwise discriminator, and therefore, Eq.~(\ref{multi}) is evidently valid.
Simulations are consistent with the results, showing that when CSDT is introduced into the population, it exhibits an evident evolutionary advantage (Figs.~\ref{fig: 4}e-j).
For instance, in populations where CSDT is absent, strategies like ALLD or WSLS may dominate, though the differences between strategies might be minimal (Figs.~\ref{fig: 4}e, g). 
However, with the introduction of CSDT, it gains a substantial evolutionary advantage (Figs.~\ref{fig: 4}f, h). 
We also plotted the proportion of strategies in a single round as a function of generation, illustrating that CSDT dominates other strategies in the later stages of evolution.

\begin{center}
\begin{figure}[h]
	\centering
	\includegraphics[width=.9\textwidth]{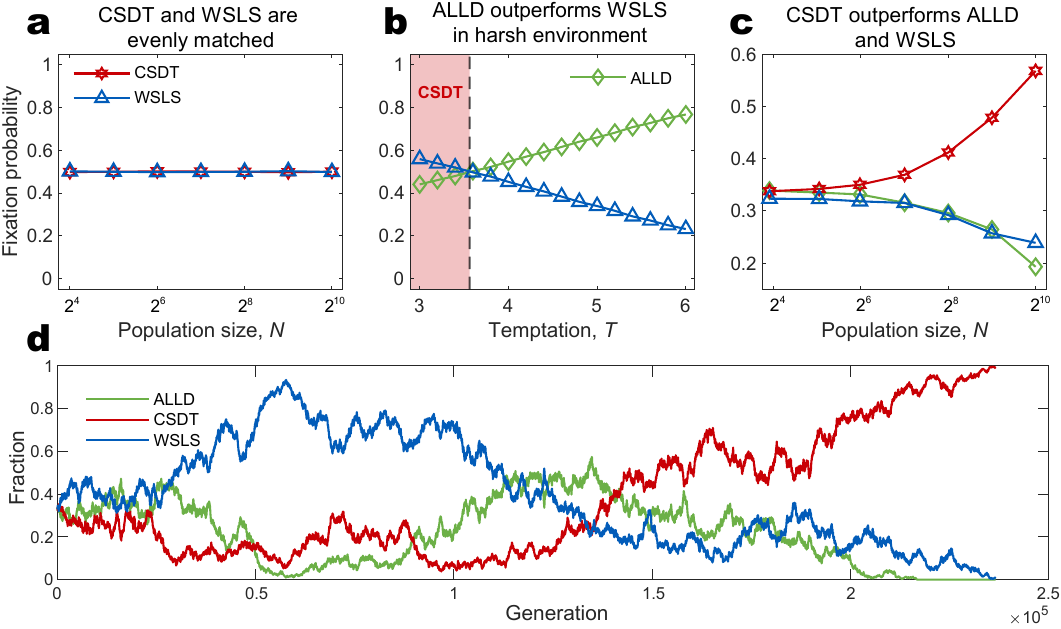}
	
	\caption{
		\textbf{Catalytic effect of ALLD.}
		\textbf{a},
		We plot the fixation probabilities of different strategies as a function of the population size $N$.
		If the population only consists of WSLS and CSDT, they are evenly matched.
		\textbf{b},
		If the environment becomes harsher, i.e., the larger temptation for defection, $T$, WSLS will not be included in the set of CSDT. And ALLD can outperform WSLS if $T$ is large.
		The vertical dash line corresponds to the critical $T^*$ for ALLD to outperform WSLS, derived by Eq.~(\ref{ResiALLD}).
		\textbf{c},
		With the inclusion of ALLD, the evolution of CSDT is promoted, suppressing WSLS and ALLD.
		\textbf{d},
		We plot the change in fractions of different strategies over generations in a single round of evolution on the network with $N=500$.
		In this round, WSLS initially has an advantage, but ALLD eventually starts to replace WSLS and is ultimately replaced by CSDT.
		Other parameters used are the same as those in Fig.~\ref{fig: 4}.
	}
	\label{fig: 5}
\end{figure}
\end{center}

However, CSDT cannot establish an evolutionary advantage over strategies that meet the other requirements of CSDT (Eqs.~(\ref{pcc1}), (\ref{pdc0}), and $p_{\text{DD}}\neq0$) but exhibit excessive tolerance; such strategies remain equally competitive with CSDT. 
Apart from CSDT, other strategies in this set fail to meet the criteria necessary to resist ALLD. 
For example, WSLS, which lacks the capability to counter ALLD in harsher environments, illustrates this issue. 
When the population comprises only CSDT and WSLS, their fixation probabilities are identical (Fig.~\ref{fig: 5}a). 
However, in harsher environments (i.e., with a higher temptation for defection $T$), WSLS no longer qualifies as a member of CSDT.
As shown in Fig.~\ref{fig: 5}b, there exists a critical value $T^*$ beyond which ALLD outperforms WSLS, a condition theoretically derived from Eq.~(\ref{ResiALLD}). 
This indicates that CSDT cannot completely shield against random drift towards strategies outside this category, leaving the population vulnerable to invasion by ALLD.

Intriguingly, the situation changes significantly when ALLD is present in the population (Fig.~\ref{fig: 5}c). 
The presence of ALLD exploits WSLS's weaknesses, rendering it unable to resist and consequently leading to an evolutionary disadvantage. 
Mathematically, this occurs because the inclusion of ALLD raises the average in Eq.~(\ref{multi}), causing the condition to be strict. 
This finding suggests that, contrary to previous research, ALLD actually facilitates the evolution of CSDT and promotes direct reciprocity. 
This effect arises from its ability to eliminate strategies that are weak to defection, thereby leaving only the strong strategies.

\section{Discussion}

Classical strategies of repeated prisoner's dilemma often focus on how to succeed in two-player games \cite{axelrod1981evolution,nowak1992TFT,press2012iterated}.
Few of them exhibit evolutionary stability, and when they do, it often requires that the payoffs for defection when facing cooperation are not excessively high \cite{nowak1992TFT,nowak1993WSLS}.
Here, we present the existence and prevalence of dominant strategies within structured populations.
The three characteristics of such strategies we derived mathematically: mutual cooperation, resistance to ALLC, and minimising the discrepancy against ALLD, provide a simple yet profound insight into how cooperation evolves. 
Such strategies always stay current actions when the opponent cooperates and chooses to tolerate defection with a limited probability, which we term ``Cooperate-Stay-Defect-Tolerate".

Despite their mathematical simplicity and clarity, our results mirror the interactions of the majority of ordinary players in the real world. 
CSDT aims for long-lasting cooperation ($p_{\text{CC}} = 1$), abhor defection, yet also tolerate it (limited $p_{\text{CD}}$, also limited but non-zero $p_{\text{DD}}$). 
And the tolerance is environment-dependent.
However, such strategies also exploit others when the opponents always cooperate ($p_{\text{DC}} = 0$). 
These characteristics are consistent with findings in sociology and psychology. 
Researches have shown that players are committed to long-term cooperation \cite{deutsch1973resolution}, and tend to punish defection, even at a cost to their own interests \cite{fehr2002altruistic,henrich2021origins}.
Simultaneously, players also exhibit a certain, albeit limited, tolerance and forgiveness \cite{mccullough2002psychology}.
In social interactions, players also exploit others to maximize their own benefits \cite{bandura1999moral,cook1987social}.
This ``folk strategy" echoes some important traits of social players.
%As we were engrossed in mathematical deduction, it seems we overlooked that the real world is a colossal evolutionary experiment. 
%Our innate instincts appear to be the result of evolution, and reality harmonizes closely with theory. 
%It might also explain why cooperation is so prevalent in the real world: because it's the strategy capable of dominating the population, which is also intrinsic to our nature.

Another interesting point to note is the role of ALLD. 
In previous studies of evolutionary dynamics, overcoming ALLD has been considered a critical indicator because of its strong invasion capability \cite{nowak1992TFT,nowak1993WSLS,nowak2006evolutionary}. 
CSDT is no exception in this regard. 
However, what sets our research apart from previous studies is the realisation that ALLD doesn't necessarily have to be entirely detrimental to cooperation. 
Instead, it can catalyze the evolution of CSDT. 
Its role is similar to that of wolves to predators: it eliminates the weaker ones, leaving behind the stronger ones \cite{mech1970wolves,ripple2004wolves}. 
Our results suggest that, under such a feedback process, defectors adjust the tolerance level in society to a safe range, preventing the population from being invaded by defectors themselves.

We have examined the influence of network structure on the feasible region of CSDT. 
Generally, networks tend to promote cooperation compared to the well-mixed populations that have been extensively studied before \cite{nowak1993WSLS,nowak2006evolutionary,hilbe2013evolution,hilbe2014extortion,hilbe2018partners,stewart2013from}. 
Structured populations allow CSDT to have higher levels of tolerance. 
An interesting direction for exploration is extending population structures to higher-order and temporal networks \cite{alvarez2021evolutionary,wang2023temporal,li2020evolution}, which could have a significant impact on the tolerance capabilities of strategies. 
Another factor that may influence tolerance is evolutionary dynamics.
Note that although we use DB updating here, the structure coefficient applies to other update rules under weak selection as well.
Generally, other update rules tend to suppress cooperation than DB updating \cite{wang2023imitation,ohtsuki2006simple}, implying a harsher environment.
It is also important to investigate whether CSDT can still dominate under strong selection \cite{hilbe2014extortion,hilbe2018partners,nowak1993WSLS}, which may be quite complex, as strong selection often exhibits great nonlinearity \cite{ibsen2015computational}.
CSDT is a simple and clear dominant strategy in networked populations, providing a fresh insight into how cooperation evolves and prevails, which remains a question that still requires further exploration.

\section{Acknowledgment}
We thank Prof. Naoki Masuda for valuable comments and suggestions.

\bibliography{ref}

\clearpage
\vspace*{\fill}
\vspace*{\fill}

\clearpage
\vspace*{\fill}
\vspace*{\fill}

\clearpage
\vspace*{\fill}
\vspace*{\fill}

\clearpage
\vspace*{\fill}
\vspace*{\fill}

\clearpage
\vspace*{\fill}
\vspace*{\fill}

\end{document}